\documentclass[letterpaper,aps,prl,preprintnumbers,twocolumn,floats,floatfix,amssymb,amsfonts,amsmath]{revtex4}
\usepackage{pgf}
\usepackage[utf8x]{inputenc}
\usepackage{aas_macros}
\usepackage{todonotes}

\renewcommand{\vec}[1]{{\mathbf{#1}}}

\usepackage{ulem}
\newcommand{\stkout}[1]{\ifmmode\text{\sout{\ensuremath{#1}}}\else\sout{#1}\fi}
\begin{document}

\preprint{TTK-19-34}

\title{Time-dependent AMS-02 electron-positron fluxes in an \emph{extended} force-field model}

\author{Marco Kuhlen}
\author{Philipp Mertsch}
\affiliation{Institute for Theoretical Particle Physics and Cosmology (TTK), RWTH Aachen University, 52056 Aachen, Germany}

\begin{abstract}The magnetised solar wind modulates the Galactic cosmic ray flux in the heliosphere up to rigidities as high as 40 GeV. In this work, we present a new and straightforward extension of the popular, but limited force-field model, thus providing a fast and robust method for phenomenological studies of Galactic cosmic rays. Our semi-analytical approach takes into account charge-sign dependent modulation due to drifts in the heliospheric magnetic field and has been validated via comparison to a fully numerical code. Our model nicely reproduces the time-dependent AMS-02 measurements and we find the strength of diffusion and drifts to be strongly correlated with the heliospheric tilt angle and magnitude of the magnetic field. We are able to predict the electron and positron fluxes beyond the range for which measurements by AMS-02 have been presented. We have made an example script for the semi-analytical model publicly available and we urge the community to adopt this approach for phenomenological studies.
\end{abstract}

\maketitle

\textit{Introduction.}---Upon entering the heliosphere, Galactic cosmic rays encounter the magnetised solar wind and are thus subject to a number of transport processes: advection with the wind, diffusion in the small-scale turbulent magnetic field, drifts due to variations of the large-scale field and adiabatic energy losses in the expanding flow. Together, these effects suppress the fluxes of cosmic rays at Earth compared to the interstellar fluxes. Collectively this is referred to as solar modulation. (See Ref.~\cite{Potgieter:2013pdj} for a review.)

For the modelling of solar modulation, two approaches have been adopted in the literature: Numerical codes solve the transport equation for models of the heliosphere of varying sophistication~\cite{Kappl:2016gsz,Vittino:2017fuh,2018AdSpR..62.2859B,2019ApJ...873...70A} and have been successfully applied to time-dependent data, too (e.g.~\cite{Potgieter2015,2019arXiv190307501J,2019ApJ...871..253C}). While such approaches have the potential to reproduce observations elsewhere in the heliosphere and thus provide a more global picture, the complexity comes at the price of a large number of unknown parameters. These parameters need to be determined by fitting the models to various observables. However, as the input interstellar fluxes depend also on unknown parameters, running such global fits is prohibitively expensive.

Phenomenological studies of Galactic cosmic ray transport on the other hand oftentimes employ the classic force-field model of Gleeson and Axford~\cite{Gleeson}. This model is conceptually simple and all the complexity of the heliosphere is condensed into only one parameter, the Fisk potential, which can be easily determined by fitting to data. In addition, allowing for this electro-static potential to be time-dependent, some degree of correlation with solar activity can be found. On the downside, the force-field model assumes a higher degree of symmetry and ignores transport processes that must be important, in particular drifts. Most importantly, the force-field model has trouble reproducing the measured fluxes. One example is crossing of fluxes, e.g.\ the proton fluxes measured by AMS-02 during Bartels rotations~\footnote{A Bartels rotation is defined as a period of 27 days, corresponding to approximately one solar rotation, starting on February 8, 1832.} 2460 and 2476 cross at $4 \, \text{GeV}$~\cite{Cuoco:2019kuu}. In the force-field model, fluxes modulated with different potentials differ at all energies and never cross.

A number of authors have tried to allow for more freedom while maintaining the simplicity of the force-field model, for instance by making the force-field potential rigidity-dependent~\cite{Cholis:2015gna,Gieseler:2017xry}. While interesting, these approaches are phenomenological fixes and have not been shown to derive from more fundamental principles, for instance from a transport equation. Here, we follow a different approach and present a systematic extension of the conventional force-field model. We start from the general transport equation in 2D that includes drifts and then reduce it--under a limited number of assumptions--to a force-field like structure. Our semi-analytical model allows calculating the modulated fluxes of any cosmic ray species at a very moderate computational cost. Our model contains two free parameters per time interval which we determine by fitting to the AMS-02 data. We further investigate temporal correlations with solar wind parameters and replace the free parameters by a linear model of tilt angle and magnetic field strength. Combining the semi-analytical model with the solar wind correlations allows reproducing the AMS-02 data and predicting electron and positron fluxes beyond the range of measurements by AMS-02.

Given its simplicity and speed, our semi-analytical model is bridging the gap between the oversimplified classical force-field model and the computationally expensive and unwieldy numerical codes. Therefore, it is ideally suited for fast likelihood evaluations and we urge the community to adopt it for phenomenological studies. We have made an example script for the semi-analytical model available in both Python and C++ at {\protect\url{https://git.rwth-aachen.de/kuhlenmarco/effmod-code}}.

\textit{Semi-analytical model}---The propagation of cosmic rays in the heliosphere is generally described in terms of a transport equation,
\begin{equation}
\frac{\partial f}{\partial t} + \nabla \cdot \left( C \vec{V} f - K \cdot \nabla f \right) + \frac{1}{3 p^2} \frac{\partial}{\partial p} \left( p^3 \vec{V} \cdot \nabla f \right) = q \, , \label{eq:transport}
\end{equation}
where $f$ is the cosmic ray phase space density with momenta $p$ measured in the observer frame, \mbox{$C = - (\partial f / \partial p) / 3$} is the Compton-Getting factor~\cite{1935PhRv...47..817C}, $\vec{V}$ denotes the solar wind velocity, $K$ is the diffusion tensor, the third term on the left hand side describes the adiabatic losses in the expanding solar wind and $q$ is a source term.

We assume a steady state and follow the conventional force-field approach insofar as we ignore the adiabatic energy losses in the fixed frame~\cite{2004JGRA..109.1101C}. Note that although we will be illustrating our method on electron and positron fluxes, all of the assumptions made are applicable for arbitrary charges and masses, that is including both non-relativistic and relativistic cosmic rays.
In the absence of sources the streaming $(C \vec{V} f - K \cdot \nabla f)$ is thus divergence-free and its integral over an arbitrary surface must vanish. After some manipulations~\footnote{See Supplemental Material for the detailed calculation.} this leads to a partial differential equation for $\tilde{f} = \tilde{f}(r,p)$
\begin{equation}
\frac{\partial \tilde{f}}{\partial r} + \frac{p \tilde{V}}{3 \tilde{K}_{r r}} \frac{\partial \tilde{f}}{\partial p} = \frac{\tilde{v}_{gc,r}}{\tilde{K}_{r r}} \tilde{f} \, , \label{eqn:modulation_equation}
\end{equation}
where tildes denote (weighted) polar angle averages and $\tilde{v}_{gc,r}$ is the average of the gradient curvature drift. The boundary condition is $\tilde{f}(r_{\text{max}}, p) = f_{\mathrm{LIS}}(p)$, $r_{\text{max}}$ being the radius of the heliosphere. Note that absorbing the non-trivial polar angle dependencies into the averaged quantities has significantly reduced the complexity. 
For the same reason current sheet drifts that have the same rigidity dependence as the gradient and curvature drifts can be absorbed into the drift term. In the following, we will therefore drop the subscript 'gc' and have $\tilde{v}_r$ denote a general drift term.

We can solve eq.~\eqref{eqn:modulation_equation} using the method of characteristics,
%
\begin{widetext}
\begin{align}
  \tilde{f}(r, p) \! = \! f_{\mathrm{LIS}}(p_{\mathrm{LIS}}(r, p)) \exp \left[ - \int_0^r \mathrm{d} r' \frac{\tilde{v}_r(r', p_{\mathrm{LIS}}(r', p))}{\tilde{K}_{r r}(r', p_{\mathrm{LIS}}(r', p)} \right] \, , \label{eqn:semi-analytical_model1}
\end{align}
\end{widetext}
where we have defined $p_\text{LIS}(r',p') = p_{r'\!,p'}(r_\text{max})$ with $p_{r'\!,p'}(r)$ a solution of the initial value problem
\begin{align}
\frac{\mathrm{d} p}{\mathrm{d} r} = \frac{p \tilde{V}}{3 \tilde{K}_{r r}} \, , \label{eqn:semi-analytical_model2} 
\end{align}
with $p(r') = p'$. We stress that the functional dependence of eqs.~\eqref{eqn:semi-analytical_model1} and \eqref{eqn:semi-analytical_model2} is novel and differs from the approaches of Ref.~\cite{Cholis:2015gna,Gieseler:2017xry}.

We will model only the momentum dependence of $\tilde{K}_{r r}$, $\tilde{v}_r$ and $\tilde{V}$ and absorb their normalisation as well as the effect of possible spatial dependencies into scaling factors by making the following replacements in eqs.~\eqref{eqn:semi-analytical_model1} and \eqref{eqn:semi-analytical_model2}:
\begin{equation}
\frac{\tilde{v}_r}{\tilde{K}_{r r}} \to g_2 \frac{\tilde{v}_r}{\tilde{K}_{r r}} \quad \text{and} \quad \frac{p \tilde{V}}{3 \tilde{K}_{r r}} \to g_1 \frac{p \tilde{V}}{3 \tilde{K}_{r r}} \, .
\end{equation}
The scaling factors $g_1$ and $g_2$ will be determined by fitting to data.

\textit{Validation}---In order to verify the validity of the approximations
made in deriving eqs.~\eqref{eqn:semi-analytical_model1} and
\eqref{eqn:semi-analytical_model2}, we have solved the transport
eq.~\eqref{eq:transport} directly using our own finite-difference code. For the
numerical solution, we need to specify a particular heliospheric transport
model. The model adopted is closely emulating one that has been successfully
applied to data from the PAMELA experiment~\cite{Potgieter2015} with some
simplifications~\footnote{The details of the heliospheric model are given in
the Supplemental Material , which includes
Refs.~\cite{1968CaJPS..46..937G,1971Ap&SS..11..288G,Webb1979,2013SSRv..176..299M,Langner,Schlickeiser1,1989Potgieter,Heber,2000Burger,Giacalone}.}.
Using the numerical code, we have confirmed that the adiabatic term, i.e.\ the
last term on the left-hand-side of eq.~\eqref{eq:transport}, always contributes
less than $10 \, \%$ to the transport equation~\footnote{See the Supplemental
Material.}.

For the semi-analytical model, we adopt fiducial parametrisations for the rigidity-dependencies of $\tilde{K}_{rr}$, $\tilde{v}_r$ and $\tilde{V}$ that follow the rigidity-dependencies used in the numerical code. The diffusion coefficient is modelled as a softly broken power law in rigidity $R$ (understood to be measured in GV),
\begin{equation}
\tilde{K}_{rr} = K_0R^a\left(\frac{R^c+R_k^c}{1+R_k^c}\right)^{(b-a)/c} \, ,
\end{equation}
with a normalization $K_0 = 30\,\text{AU}^2/\text{d}$~\footnote{$1 \, \text{AU}$ denotes the average distance of the Sun and Earth and equals $ 1.5 \times 10^{13} \, \text{cm}$. $1 \, \text{d}$ denotes a mean solar day and equals $86,400\,\text{s}$.}, power law indices $a = 0$ and $b = 1.55$, $c=3.5$ and a break rigidity $R_k = 0.28$ fixed to values obtained for electrons and positrons in previous studies~\cite{Potgieter2015}. The radial drift velocity can be parametrised as
\begin{equation}
\tilde{v}_r = \kappa_{\text{0}} \frac{ R}{3 B_0}\frac{10\,R^2}{1+10\,R^2} \, ,
\end{equation}
where we set $\kappa_\text{0} = 1\,\text{AU}/\text{d}$ and the magnetic field $B_0 = 1 \, \mu\text{G}$. Note that for $R \lesssim 40\,\text{GV}$ and the parameter values that we consider no pathological behaviour occurs.

The averaged radial solar wind obtains a momentum dependence due to the non-uniform spatial distribution of cosmic rays in the heliosphere which depends on the polarity cycle. Motivated by results of our numerical simulation it will be modeled as a step function in momentum
\begin{equation}
\tilde{V} = V_0(1+\Delta V\,\theta(R-R_b)) \, .
\label{eq:v}
\end{equation}
While the rigidity $R_\text{b}$ and the height of the step $\Delta V$ have to be fitted to data for different products of charge sign and magnetic field polarity, the normalization $V_0$ is degenerate with $K_0$ and $\kappa_{\text{0}}$ and has thus been fixed to $620 \, \text{km} \, \text{s}^{-1}$. We note that the momentum dependence of $\tilde{V}$ only affects energies lower than those of the AMS-02 or PAMELA electron and positron measurements.

We have validated our semi-analytical method by confirming that the fluxes modulated according to eqs.~\eqref{eqn:semi-analytical_model1} and \eqref{eqn:semi-analytical_model2} agree with the results from the fully numerical code. Our semi-analytical method always agrees with the code to within $10 \, \%$ which is far better than for the conventional force-field model~\footnote{Plots for $qA>0$ and $qA<0$ are shown in the Supplemental Material.}.

\begin{figure}
\includegraphics[width=.5\textwidth]{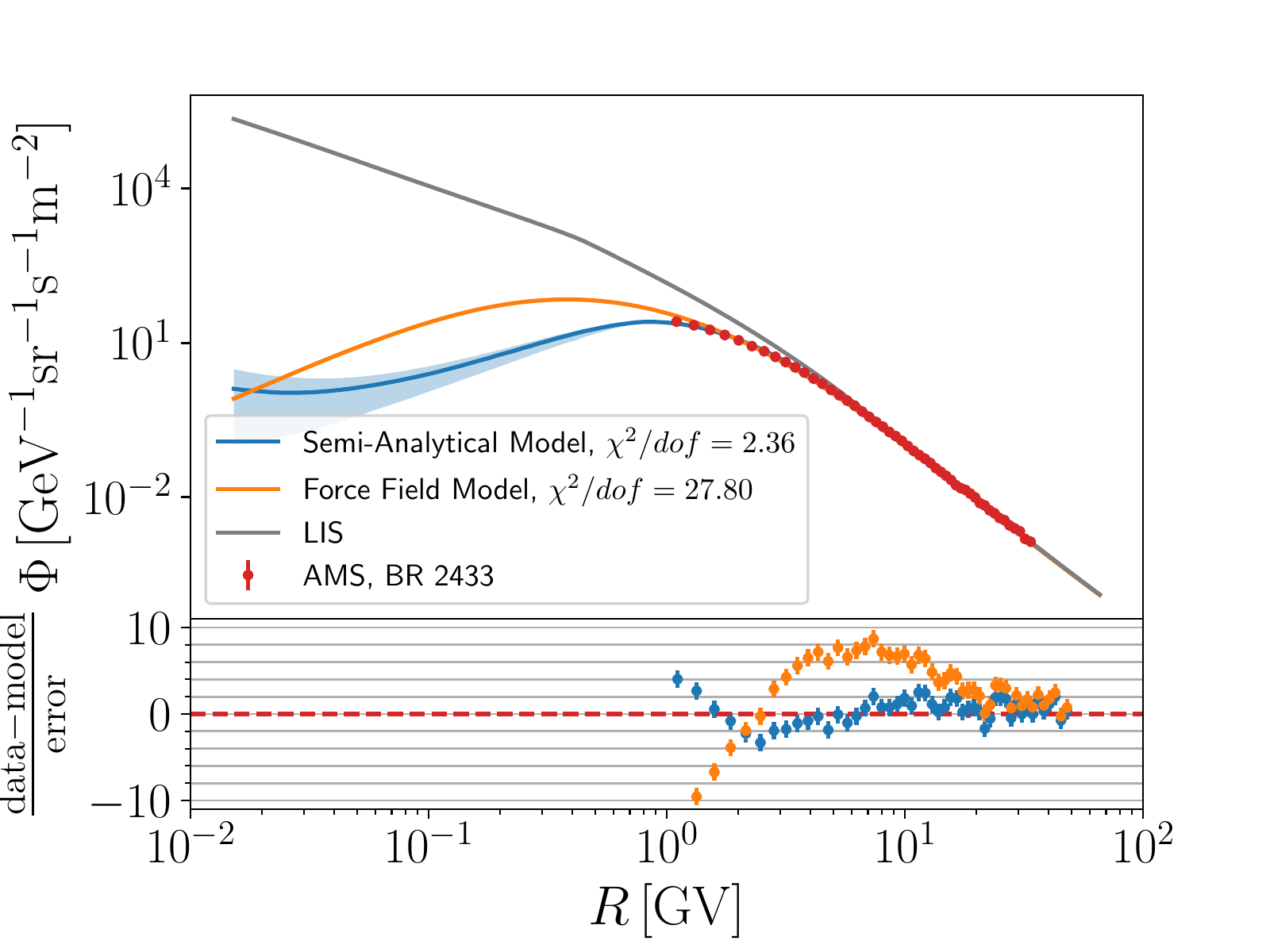} 
\includegraphics[width=.5\textwidth]{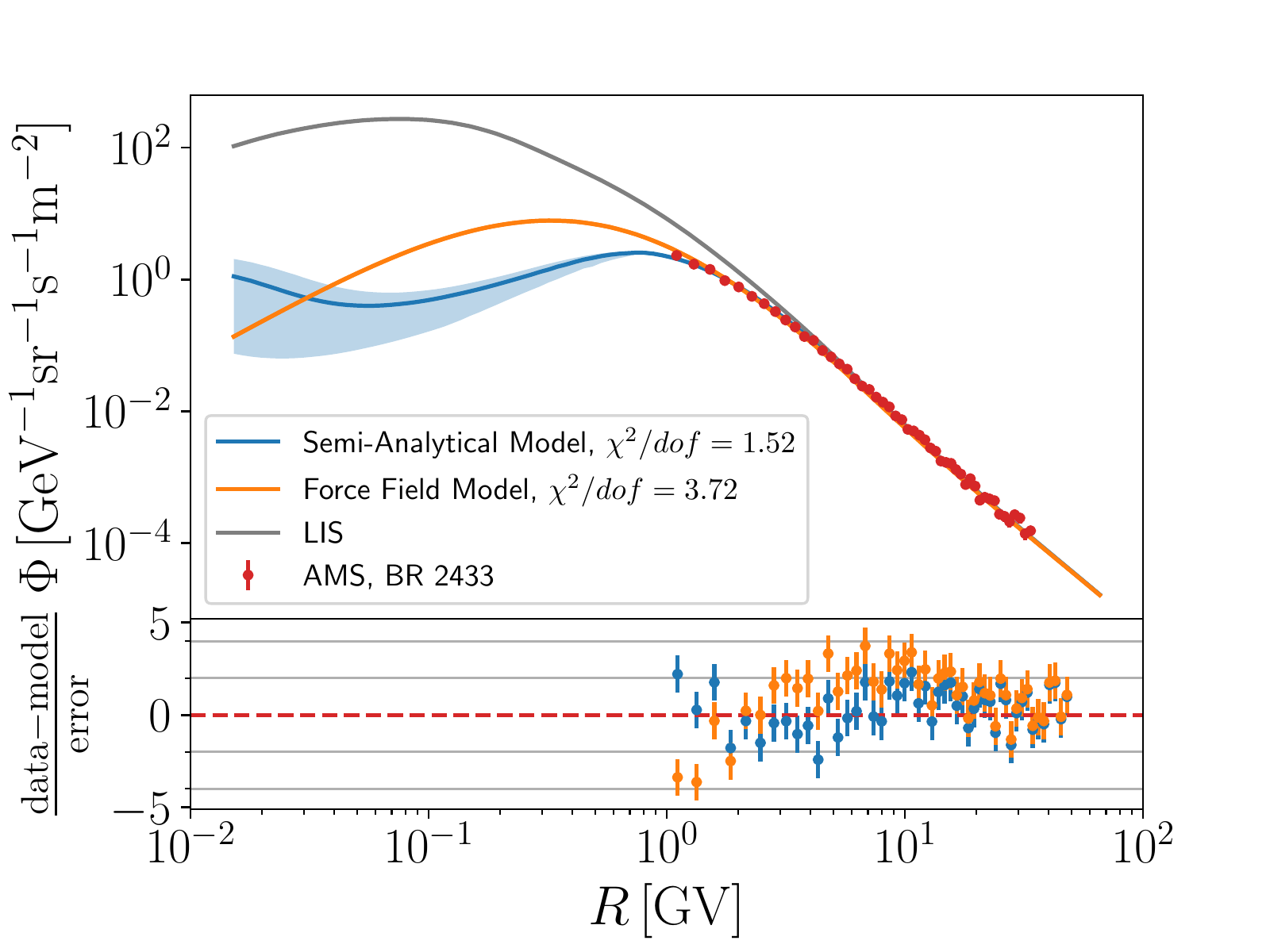}
  \caption{Fit of our semi-analytical model (blue line) to AMS-02 data~\cite{ams} and the conventional force-field fit (orange line) for Bartels rotation 2433 for electrons (upper panel) and positrons (lower panel). In both cases the local interstellar spectrum (grey line) is taken from Ref.~\cite{Vittino:2019yme}. The lower part of the plots shows the pull distribution for the fit. For the blue line the step in the solar wind is fitted while for the blue band it is varied between $\Delta V=0$ and $0.4$.
  }
\label{fig:amsfit}
\end{figure}

\textit{Application to experimental data.}---In a first step, we determine for each time bin separately the scaling parameters $g_1$ and $g_2$. The resulting modulated fluxes for one exemplary Bartels rotation is shown in Fig.~\ref{fig:amsfit} and compared to AMS-02~\cite{ams} data. We also show the results from the conventional force-field model. In particular for the case of electrons the agreement with data is much better for the semi-analytical model than for the conventional force field model. One possible reason for remaining deviations of the experimental data from the model are events of heliospheric origin, like coronal mass ejections, which are not taken into account in our model. Other reasons include inaccuracies in the interstellar fluxes adopted or too restrictive rigidity-dependencies of $\tilde{K}_{rr}$ and $\tilde{v}_r$.

\textit{Prediction of electron-positron fluxes.}---While the semi-analytical model with fitted parameters $g_1$ and $g_2$ for most time intervals nicely reproduces the data, this procedure is not predictive. In order to \emph{predict} modulated fluxes, we need to model the parameters $g_1$ and $g_2$ as a function of time. Physically, we would expect them to be (anti-)correlated with solar wind parameters. For example, $g_1$ is related to $K_0$, which parametrises the strength of diffusion, and could be correlated with a proxy for solar activity, e.g.\ the tilt angle $\alpha$, while $g_2$, which parametrises the relative strength of drifts, could be anti-correlated with the strength of the magnetic field, $B$. We stress that both the tilt angle as measured in the solar corona and the field strength as measured by ACE are relatively local observables whereas the cosmic ray particles spend a finite time in the heliosphere. We would therefore expect the fitted parameters to be affected only with a certain delay and after averaging over time.

We have therefore searched for linear correlations between the $g_1$ and $g_2$ fitted to AMS-02 data and the tilt angle and magnetic field strength using a moving average of width $\tau$, $\langle \alpha \rangle_{\tau}$ and $\langle B \rangle_{\tau}$. It is generally accepted that particles of different signs ($q$) arrive preferably along different directions in different polarity ($A$) cycles, specifically for $qA>0$ ($qA<0$) particles arrive mostly from the polar (equatorial) direction. Due to the waviness of the current sheet propagation through the equatorial region takes longer than through the polar region and this implies also different delays for $qA>0$ and $qA<0$ with electrons ($q<0$) in an $A>0$ cycle experiencing a similar, large delay as positrons ($q>0$) in an $A<0$ cycle. We therefore allow for different widths $\tau$ for different $qA$.

We find good correlations (Pearson correlation coefficients of $\sim 0.9$) between $g_1$ and $\langle \alpha \rangle_{\tau}$ for widths of $\tau = 12$ and $39$ Bartels rotations for $qA>0$ and $qA<0$, respectively. For $g_2$, we find an anti-correlation ($\sim -0.65$) with $\langle B \rangle_\tau$ for widths of $\sim 5$ Bartels rotations. We therefore model the scaling factors as $g_1^{\pm} = a_1^\pm + b_1^\pm \langle \alpha \rangle_{\tau}$ and $g_2^{\pm} = a_2^\pm + b_2^\pm \langle B \rangle_{\tau}$. To allow for a smoother transition around the solar maximum we adopt $\tau$'s of $21$ and $30$ Bartels rotations for $qA>0$ and $qA<0$ closer to the solar maximum. We summarise the time ranges and averaging widths adopted in Tbl.~\ref{tbl1}. The beginning and end of the ranges correspond to regions where the polarity could be determined including an additional delay.

\begin{table}[!b]
\resizebox{\columnwidth}{!}{%
\begin{tabular}{cccc | cc | cc | cc | cc}
\hline
start 	& end	& $\tau_{-}$ 	& $\tau_{+}$ 	& $a_1^-$ 	& $a_1^+$ 	& $b_1^-$ 	& $b_1^+$ & $a_2^-$ 	& $a_2^+$ 	& $b_2^-$ 	& $b_2^+$ \\
  \multicolumn{4}{c |}{$[ \text{BR} ]$}		& \multicolumn{2}{c |}{} & \multicolumn{2}{c |}{$[1/\text{deg}]$} & \multicolumn{2}{c |}{} & \multicolumn{2}{c}{$[1/\mu\text{G}]$}  \\
\hline\hline
  2239 	& 2435	& 12				& 39				& 1.79 & -0.32 & 0.025 &0.082 & 1.50&5.80 &-0.441 & -1.30\\
2435 	& 2460	& 21				& 30			& -2.77& -0.88 & 0.121& 0.071& 7.59& 7.34& -1.62& -1.57\\
2460 	& 2485	& 30				& 21				& 5.14& -0.46 & 0.012& 0.068& 6.18& 5.46& -1.31 & -1.16\\
2485	& 2530	& 39				& 12				& 0.37& -1.57 & 0.094& 0.107& -0.049& 3.95& -0.313& -0.808\\
\hline
\end{tabular}
}
\caption{Interval ranges and averaging widths for electrons ($\tau_{-}$) and positrons ($\tau_{+}$) as well as model coefficients.}
\label{tbl1}
\end{table}

\begin{figure*}
\includegraphics[width=0.9\textwidth]{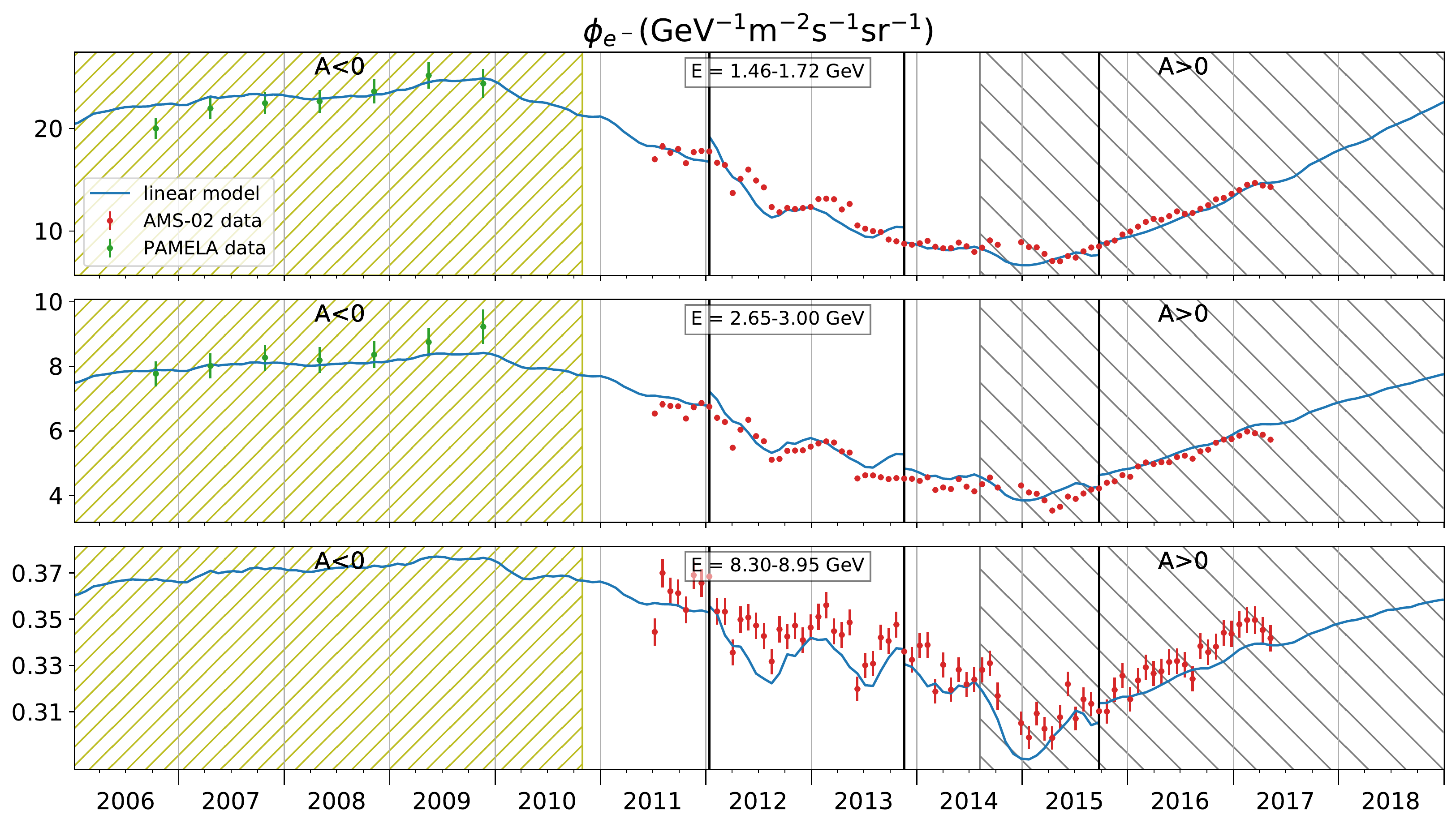}
\includegraphics[width=0.9\textwidth]{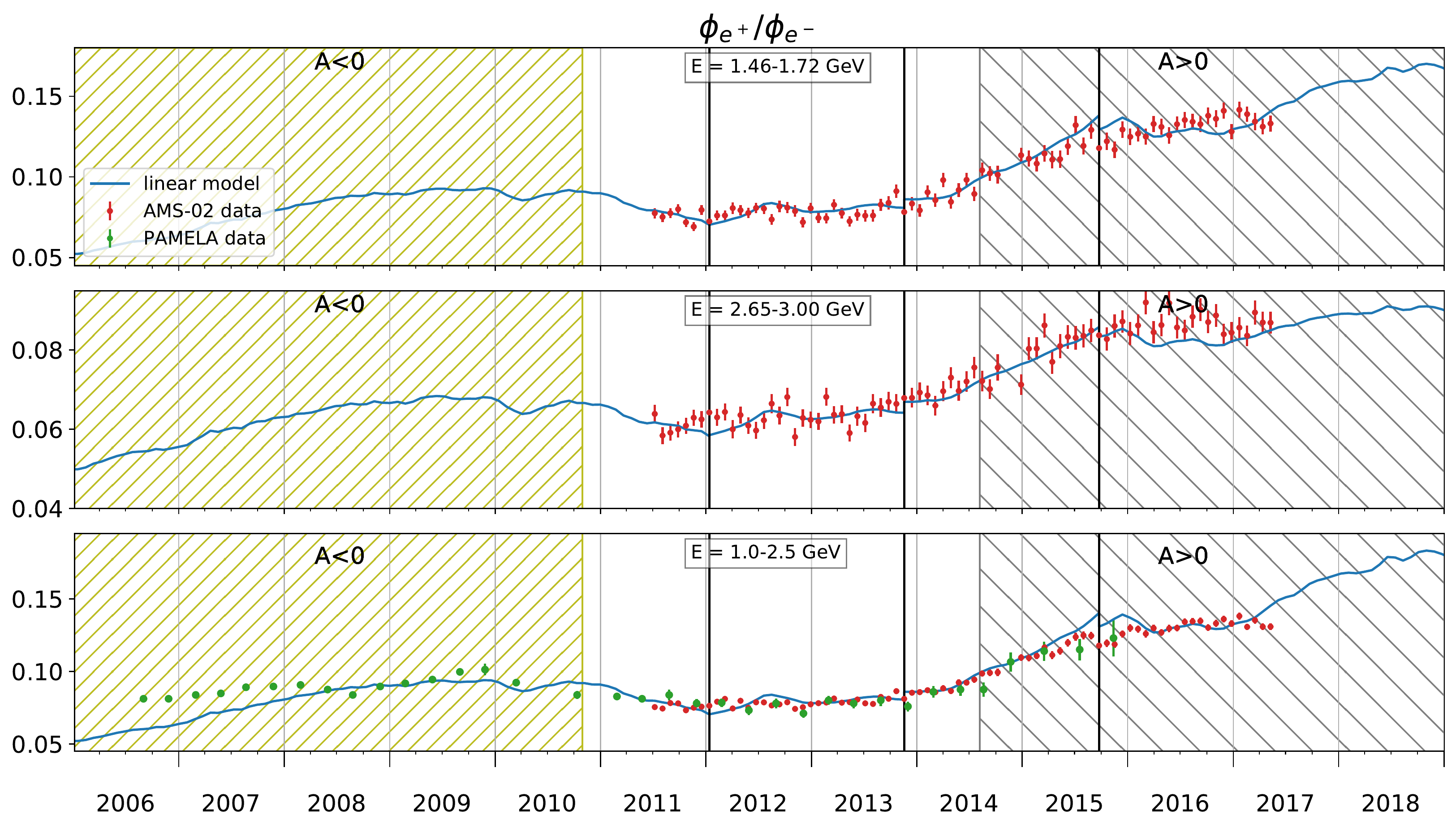}
\caption{Time-dependent electron flux (upper panels) and positron ratio (lower panels), calculated with our linear model (blue line) and compared to AMS-02 data (red dots). The bottom panel shows an average over multiple bins in order to compare to PAMELA data~\cite{Munini:2017ali} (green dots) with wider energy bins. The vertical lines show the transitions between different ranges around the solar maximum. The hatched regions before 2011 and starting in 2014 denote the periods for which the polarity could be unambiguously determined~\cite{polarity}.} 
\label{fig:predictedfluxes}
\end{figure*}

In Fig.~\ref{fig:predictedfluxes}, we show the electron fluxes and positron-to-electron ratios as a function of time for different energy ranges. There are some deviations from the data that take place on short time scales and are usually coincident with known Forbush decreases that cannot be captured by our model. In order to assess the quality of the fit, one needs to compare against an alternative model for the long term variation. A harmonic fit of the fluxes in each of the $49$ individual energy bins $i$ as $\phi_i = A_i + B_i \cos(2 \pi t/T_i + \varphi_i)$ seems as agnostic a model as possible. In this harmonic model with $49 \times 4$ free parameters, the $\chi^2 / \text{dof}$ is $35920/3626$ for electrons and $19447/3626$ for positrons. This is to be compared to $26384/3806$ for electrons and $8028/3806$ for positrons in our model ($4$ time intervals with $4$ parameters each). In the positron ratio the short term features cancel and the  $\chi^2/\text{dof} = 4991/3790$ is much better than for the electron flux or the positron flux alone.

While our model thus reproduces the measurements by AMS-02 rather well, we can predict the fluxes and the ratio also at other times. First, our model can be straight-forwardly extended to earlier times. We note that we also reproduce the measurements of PAMELA even though PAMELA data have not been used in the fitting of the coefficients $a^{\pm}_1$, $b^{\pm}_1$, $a^{\pm}_2$ or $b^{\pm}_2$. Second, we can also make a prediction for the fluxes and the ratio for the range for which measurements of $\alpha$ and $B$ are available, in principle until the end of the $A>0$ polarity cycle. As a general trend, we would expect the tilt angle to start rising again soon, leading to a falling positron ratio and subsequent maxima in the positron and electron fluxes.

\textit{Conclusion.}---We have presented a 2D semi-analytical model that incorporates charge-sign dependent modulation effects due to drifts in the solar magnetic field. We have validated our model by comparing to the results of our finite-difference code and found that the semi-analytical model is significantly more accurate than the conventional force field model. Our model has four free parameters, two of which encode the strength of diffusion and drifts.

It is remarkable that allowing these two free parameters to vary per time interval we can fit the time-dependent AMS-02 electron and positron fluxes rather accurately. Introducing in addition a linear model relating these two parameters to solar wind parameters, we can reproduce not only the AMS-02 measurements, but also the PAMELA electrons fluxes which do not overlap with the AMS-02 electron fluxes. We can also predict the electron and positron fluxes beyond the range for which data has been published. We stress that our method is significantly faster than numerical methods and more accurate than the standard force field model. It can thus be applied to large datasets with fine time binning like the current AMS-02 data. Given its simplicity, it should therefore be applied in phenomenological studies of cosmic ray transport and we facilitate such applications by providing an example script at {\protect\url{https://git.rwth-aachen.de/kuhlenmarco/effmod-code}}.

\textit{Acknowledgments.}---We are grateful to Henning Gast, Stefan Schael and Andrea Vittino for helpful discussions.

\bibliographystyle{unsrtnat}
\bibliography{Bibliography}

\end{document}


\title{Time-dependent AMS-02 electron-positron fluxes in an \emph{extended} force-field model: Supplementary Material}

\author{Marco Kuhlen}
\author{Philipp Mertsch}

\affiliation{Institute for Theoretical Particle Physics and Cosmology (TTK), RWTH Aachen University, 52056 Aachen, Germany}

\maketitle

\section{Conventional Force Field Model}

The conventional force-field model~\cite{Gleeson,1971Ap&SS..11..288G} is most easily explained by starting with the cosmic ray transport equation in the form of Gleeson \& Webb (cf.\ e.g.~\cite{2004JGRA..109.1101C}),
\begin{equation}
\frac{\partial f}{\partial t} + \nabla \cdot \left( C \vec{V} f - K \cdot \nabla f \right) + \frac{1}{3 p^2} \frac{\partial}{\partial p} \left( p^3 \vec{V} \cdot \nabla f \right) = q \, . \label{eqn:transport}
\end{equation}
Here, $f$ is the cosmic ray phase space density in the observer's frame, $C = - (\partial f / \partial p) / 3$ is the Compton-Getting factor, $\vec{V}$ denotes the solar wind, $K$ is the diffusion tensor and the third term on the left hand side describes the adiabatic losses in the expanding solar wind.

Assuming a steady state $\partial f/\partial t = 0$, no sources $q = 0$ and neglecting the adiabatic momentum loss rate in the observer's frame, \mbox{$\langle\dot p\rangle = \frac{1}{3}\,\mathbf{V}\,\cdot\,\nabla\,f/f = 0$}, this reduces to
\begin{equation}
  \nabla\cdot\left(C\mathbf{V}f - \mathbf{K}\cdot\nabla f= 0\right) \, . \label{eqn:zero_streaming}
\end{equation}
Note that this does not imply that individual particles do not lose energy. $\langle\dot p\rangle = \frac{1}{3}\,\mathbf{V}\,\cdot\,\nabla\,f/f$ is the average rate of momentum change \emph{in the observer's frame} while the mean rate of change of momentum of an individual particle in the solar wind frame is $\langle\dot p\rangle = -\frac{p}{3} \nabla \cdot\mathbf{V}$~\cite{Webb1979}.

Assuming spherical symmetry, eq.~\eqref{eqn:zero_streaming} further reduces to
\begin{equation}
CVf - \kappa\frac{\partial f}{\partial r } = 0 \, ,
\label{eq:zerostreaming}
\end{equation}
where obviously all structure of the heliospheric magnetic field, anisotropic diffusion and any drift terms are lost. Eq.~\eqref{eq:zerostreaming} is also called the zero streaming condition and indeed in an approximate one dimensional solution of the full transport equation it was shown that the advective and diffusive flow almost cancel~\cite{1968CaJPS..46..937G}.

Substituting the Compton-Getting factor gives the first order partial differential equation,
\begin{equation}
\frac{\partial f}{\partial r} + \frac{Vp}{3\kappa}\frac{\partial f}{\partial p} = 0 \, ,
\label{eq:forcefield}
\end{equation}
which can now be solved using the method of characteristics, that is $f$ is constant along the characteristics,
\begin{equation}
\frac{\partial p}{\partial r} = \frac{Vp}{3\kappa} \, .
\label{eqn:characteristics}
\end{equation}

If the diffusion coefficient factorises into a radial and a momentum dependence, $\kappa = \beta \kappa_r(r) \kappa_p(p)$, eq.~\eqref{eqn:characteristics} can easily be integrated to
\begin{equation}
\int_{p_\text{TOA}}^{p_\text{LIS}} \frac{\beta \kappa_{p'}}{p'} dp' = \int_{r_\text{TOA}}^{r_\text{LIS}}\frac{V}{3\kappa_r(r')}dr'\equiv \phi(r),
\end{equation}
where the indices TOA and LIS denote the top of the atmosphere and the local interstellar spectrum respectively.

When further assuming $\kappa_p \propto p$ and $\beta \approx 1$ for relativistic particles the force-field $\phi(r)$ reduces to the well-known form $\phi = p_{LIS} - p_{TOA}$. 

The conservation of the phase-space density $f$ along characteristics, cf.\ eq.~\eqref{eq:forcefield}, implies (again for relativistic particles) the following relation between the interstellar (unmodulated) flux $J_{LIS}$ and the top of the atmosphere (modulated) flux $J_{TOA}$
\begin{equation}
\frac{J_{TOA}}{p^2_{TOA}} = \frac{J_{LIS}}{p^2_{LIS}} \, .
\end{equation}

\section{Extended force field model}

We start by assuming a steady state and follow the conventional force-field approach by neglecting the adiabatic energy losses in the observer's frame~\cite{2013SSRv..176..299M}. Below, we will show that the adiabatic term always contributes less than 10\% to the transport equation (cf. Sec.~``Validation'' and Fig.~\ref{fig:adiabatic}). In the absence of sources the streaming $\vec{F} \equiv C \vec{V} f - K \cdot \nabla f$ is thus divergence-free and its integral over an arbitrary surface must vanish. We choose this surface to be a sphere of radius $r$ such that the divergence-free streaming condition in spherical coordinates, $\{r, \theta, \phi \}$, reads
\begin{equation}
\int_0^{\pi/2} \mathrm{d} \theta \sin \theta \left( C V f - \left( K_{r r} \frac{\partial f}{\partial r} + K_{r \theta} \frac{1}{r} \frac{\partial f}{\partial \theta} \right) \right) = 0 \, .
\end{equation}
 
For the last term, we apply integration by parts,
\begin{equation}
\int_0^{\pi/2} \mathrm{d} \theta \sin \theta \left(  C V f  -  K_{r r} \frac{\partial f}{\partial r}\right) + \frac{1}{r} \int_0^{\pi/2} \mathrm{d} \theta f \frac{\partial}{\partial_{\theta}} \left( \sin\theta K_{r \theta}  \right) \nonumber - \frac{1}{r} \left[ \sin \theta K_{r \theta} f \right]_0^{\pi/2} = 0 \, .
\end{equation}

Next, we identify the Compton-Getting factor $C = - (\partial \ln f / \partial \ln p) / 3$, the off-diagonal term of the diffusion tensor $K_{r \theta} = -\kappa_A \sin \psi$ ($\psi$ denoting the spiral angle of the magnetic field) and the radial component of the drift velocity $\partial_{\theta} \left(\sin \theta K_{r \theta} \right) / (r \sin \theta) = v_{gc,r}$ leading to
\begin{equation}
\int_0^{\pi/2} \mathrm{d} \theta \sin \theta \left( - \frac{pV}{3} \frac{\partial f}{\partial p} -  K_{r r} \frac{\partial f}{\partial r} \right) + \int_0^{\pi/2} \mathrm{d} \theta \, \sin \theta \, v_{gc,r} f  + \frac{1}{r} \left[ \sin \theta \kappa_A \sin\psi f \right]_0^{\pi/2} = 0 \, .
\end{equation}
We note that under the assumption that $f(\theta = \pi/2) \approx \tilde{f}$, the last term can be absorbed into the previous one (that also contains the drift velocity) if their momentum dependences are identical. This assumption has been verified using the numerical solution of the transport equation. We find good agreement between the third term with $f(\theta = \pi/2)$ and the same term with $\tilde{f}$ in all cases where this term gives a meaningful
contribution to this equation. 

Introducing the angular averages,
\begin{align}
\tilde{f} &= \int_0^{\pi/2} \mathrm{d} \theta \sin\theta f \, ,
\\ \tilde{V} &= \left(\partial_p \tilde{f} \right)^{-1} \int_0^{\pi/2} \mathrm{d} \theta \sin\theta V \partial_p f \, ,
\\ \tilde{K}_{r r} &= \left(\partial_r \tilde{f} \right)^{-1} \int_0^{\pi/2} \mathrm{d} \theta \sin\theta K_{r r} \partial_r f \, ,
\\ \tilde{v}_{gc,r} &= \left( \tilde{f} \right)^{-1} \int_0^{\pi/2} \mathrm{d} \theta \, \sin \theta \, v_{gc,r} f \, ,
\end{align}
this leads to the partial differential equation
 \begin{equation}
\frac{\partial \tilde{f}}{\partial r} + \frac{p \tilde{V}}{3 \tilde{K}_{r r}}                      \frac{\partial \tilde{f}}{\partial p} =  \frac{\tilde{v}_{gc,r}}{\tilde{K}_{r r}} \tilde{f} \, . \label{eqn:modulation_equation}
\end{equation}

We can solve eq.~\eqref{eqn:modulation_equation} using the method of characteristics,
\begin{align}
  \tilde{f}(r, p) \! = \! f_{\mathrm{LIS}}(p_{\mathrm{LIS}}(r, p)) \mathrm{e}^{ - \int_0^r \mathrm{d} r' \frac{\tilde{v}_{gc,r}(r', p_{\mathrm{LIS}}(r', p))}{\tilde{K}_{r r}(r', p_{\mathrm{LIS}}(r', p)} } \, , \label{eqn:semi-analytical_model1}
\end{align}
with the definition $p_\text{LIS}(r',p') = p_{r'\!,p'}(r_\text{max})$ and where $p_{r'\!,p'}(r)$ is a solution of the initial value problem
\begin{align}
\frac{\mathrm{d} p}{\mathrm{d} r} = \frac{p \tilde{V}}{3 \tilde{K}_{r r}} \, , \label{eqn:semi-analytical_model2} 
\end{align}
with $p(r') = p'$.

\section{Validation}
\label{sec:validation}

In order to verify the validity of the approximations made in deriving eqs.~\eqref{eqn:semi-analytical_model1} and \eqref{eqn:semi-analytical_model2}, we have solved the transport eq.~\eqref{eqn:transport} directly using our own 2D finite-difference code that is similar in structure to the one by Langer~\cite{Langner}. Below, we specify the particular heliospheric transport model adopted which is closely emulating a model that has been successfully applied to data from the PAMELA experiment~\cite{Potgieter2015}. Where necessary, we focus on propagation parameters relevant for relativistic electrons and positrons for application to the recently presented time-dependent measurements of the electron and positron fluxes from AMS-02.

The parametrisation of the radial solar wind is taken as a product of two functions $V_r$ and $V_\theta$ that only depend on the radial coordinate and the latitude respectively:
\begin{equation}
\begin{aligned}
V_r &= V_0 \left(1-\exp\left(\frac{40}{3r_0}(r_\odot-r)\right)\right) \\
V_\theta &= 1.5 - 0.5\cdot\tanh\left(16(\theta-\frac{\pi}{2}+\phi)\right) \, ,
\end{aligned}
\end{equation}
where $r_0 = 1~\text{AU}$ is the distance from the Earth to the Sun, $r_\odot = 0.005~\text{AU}$ is the radius of the Sun, $V_0 = 400 \, \text{km} \, \text{s}^{-1}$ and $\phi = \alpha + 15\pi/180$. Therefore the solar wind varies between $\sim 400  \, \text{km} \, \text{s}^{-1}$ in the equatorial region and $\sim 800  \, \text{km} \, \text{s}^{-1}$ in the polar region with the transition depending on the tilt angle $\alpha$.

In all generality, the diffusion tensor can be written as
\begin{equation} 
  K = 
  \begin{bmatrix}
  \kappa_\parallel & 0 & 0 \\
  0 & \kappa_{\perp\theta} & \kappa_A \\
  0 & -\kappa_A & \kappa_{\perp r}
\end{bmatrix}
  = K_S + K_A,
  \label{eq:difftensor}
\end{equation}
in a coordinate system with the first coordinate axis aligned with the background magnetic field direction. Here $\kappa_\parallel$ is the diffusion coefficient parallel to the background magnetic field and $\kappa_{\perp\theta}$ and $\kappa_{\perp r}$ are the diffusion coefficients perpendicular to the magnetic field in polar and radial direction, respectively.

Assuming a Parker-like magnetic field the diffusion tensor can be transformed from coordinates aligned with the magnetic field to the corotating frame as
\begin{equation}
    \begin{bmatrix}
    K_{rr} & K_{r\theta} & K_{r\phi} \\
    K_{\theta r} & K_{\theta\theta} & K_{\theta\phi} \\
    K_{\phi r} & K_{\phi\theta} & K_{\phi\phi} \\
  \end{bmatrix}
     = \begin{bmatrix}
      \kappa_\parallel \cos^2\psi + \kappa_{\perp r}\sin^2\psi & -\kappa_A\sin\psi & (\kappa_{\perp r} - \kappa_\parallel)\cos\psi\sin\psi \\
      \kappa_A\sin\psi & \kappa_{\perp\theta} & \kappa_A\cos\psi \\
      (\kappa_{\perp r} - \kappa_\parallel)\cos\psi\sin\psi &- \kappa_A\cos\psi & \kappa_\parallel\sin^2\psi+\kappa_{\perp r} \cos^\psi
    \end{bmatrix},
\end{equation}
with the $\psi$ the spiral angle of the magnetic field.

The parametrisation of the transport parameters is based on the model described in~\cite{Potgieter2015}. The parallel diffusion coefficient is parametrised as a softly broken power law,
\begin{equation}
\kappa_\parallel = \kappa_\parallel^0 \beta \left(  \frac{B_0}{B}  \right) \left( R/R_0 \right)^a \!\left[\frac{\left(R / R_0\right)^c+\left(R_k / R_0\right)^c}{1+\left(R_k / R_0\right)^c}\right]^{(b-a)/c} \!\!\!
\end{equation}
Here, $R$ is the particle rigidity defined as momentum over particle charge and $\beta=v/c$, $B_0 = 1\,\text{nT}$ and $R_0 = 1\,\text{GV}$ are reference values for the magnetic field strength and momentum, $R_k$ is the break rigidity and $\kappa_\parallel^0$ is a normalisation factor of the order of $\sim 30\,\text{AU}^2 \, \text{d}^{-1}$. For $R/R_k \ll 1$, $\kappa_\parallel\propto R^a$, whereas for $R/R_k \gg 1$, $\kappa_\parallel\propto R^b$ and the parameter c determines the softness of the break. The parameters of the diffusion coefficient can be determined by fitting the model to experimental data. Throughout this work we set $a=0$ which is consistent with theoretical calculations of the mean free path of electrons~\cite{Schlickeiser1}. For simplicity the diffusion coefficients perpendicular to the magnetic field have the same energy dependence and are only rescaled by a factor $f_\perp = 0.02$.

The term of the transport equation involving the anti-symmetric part of the diffusion tensor $K_A$ can be rewritten as an additional advection term with the drift velocity 
\begin{equation}
\mathbf{V_d} = \nabla \times (\kappa_A \mathbf{e_B}) \, ,
\end{equation}
where $\mathbf{e_B}$ is a unit vector pointing in the direction of the regular magnetic field. Under the assumption of weak scattering the drift coefficient is
\begin{equation}
\kappa_A = \kappa_A^0 \frac{v (R/R_0)}{3 cB},
\end{equation}
with $\kappa_A^0$ a dimensionless constant. The cases $\kappa_A^0 = 1$, $0.5$ and $0$  have been referred to as refer to as ``full drifts'', ``half drifts'' and ``no drifts'', respectively~\cite{1989Potgieter}.

The weak scattering form of $\kappa_A$ cannot be used to explain the small latitudinal gradients observed by Ulysses~\cite{Heber}, but the modification,
\begin{equation}
\kappa_A = \kappa_A^0 \frac{v (R/R_0)}{3 c B}\left(\frac{10 (R/R_0)^2}{1+10 (R/R_0)^2}\right),
\end{equation}
leads to better agreement with data since it reduces drifts for particles with very low energies slightly~\cite{2000Burger}.  This choice for $\kappa_A$ is also consistent with numerical simulations done by Giacalone et al.~\cite{Giacalone}.

\begin{figure}
\includegraphics[width=0.5\textwidth]{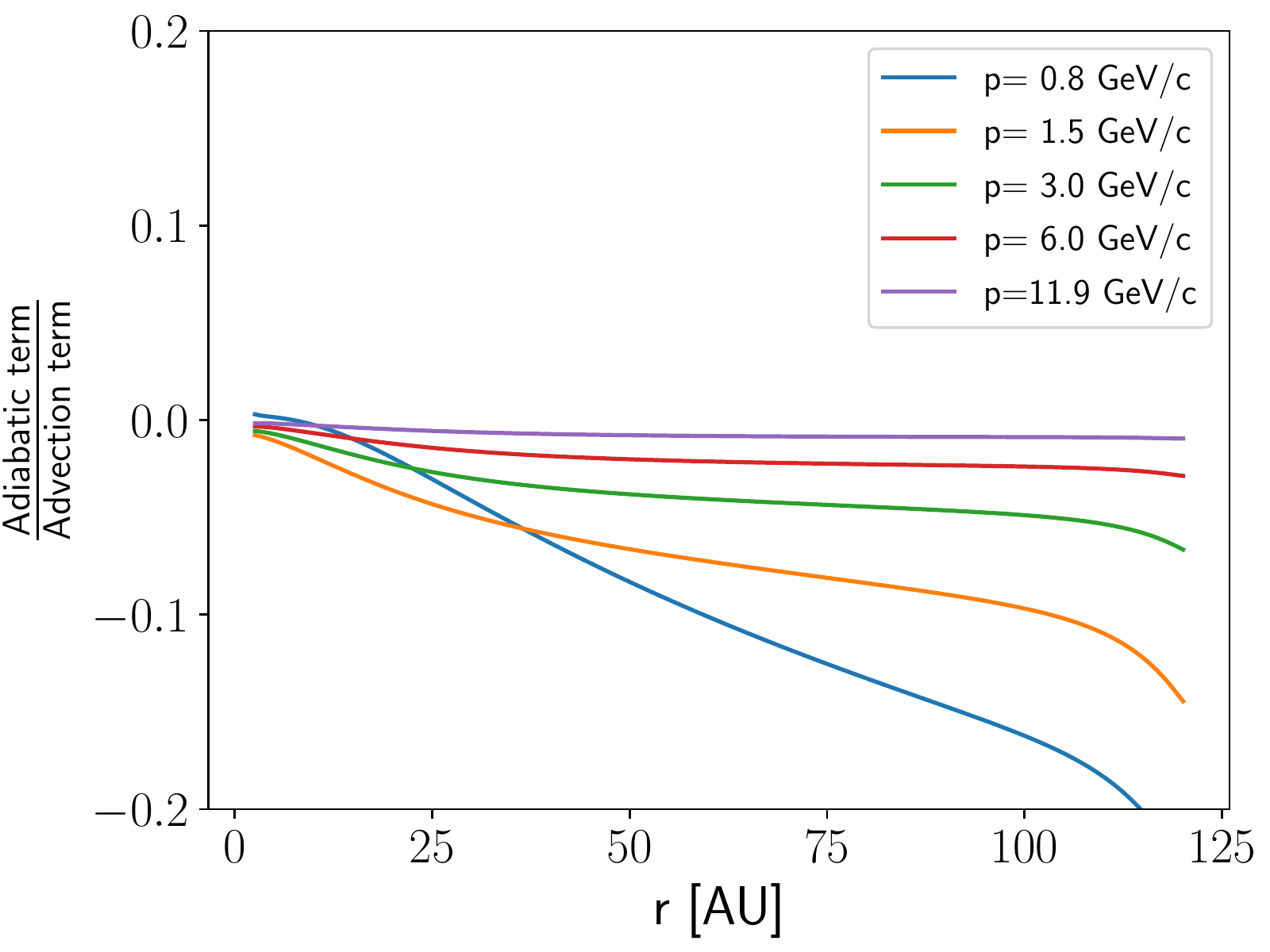}\includegraphics[width=0.5\textwidth]{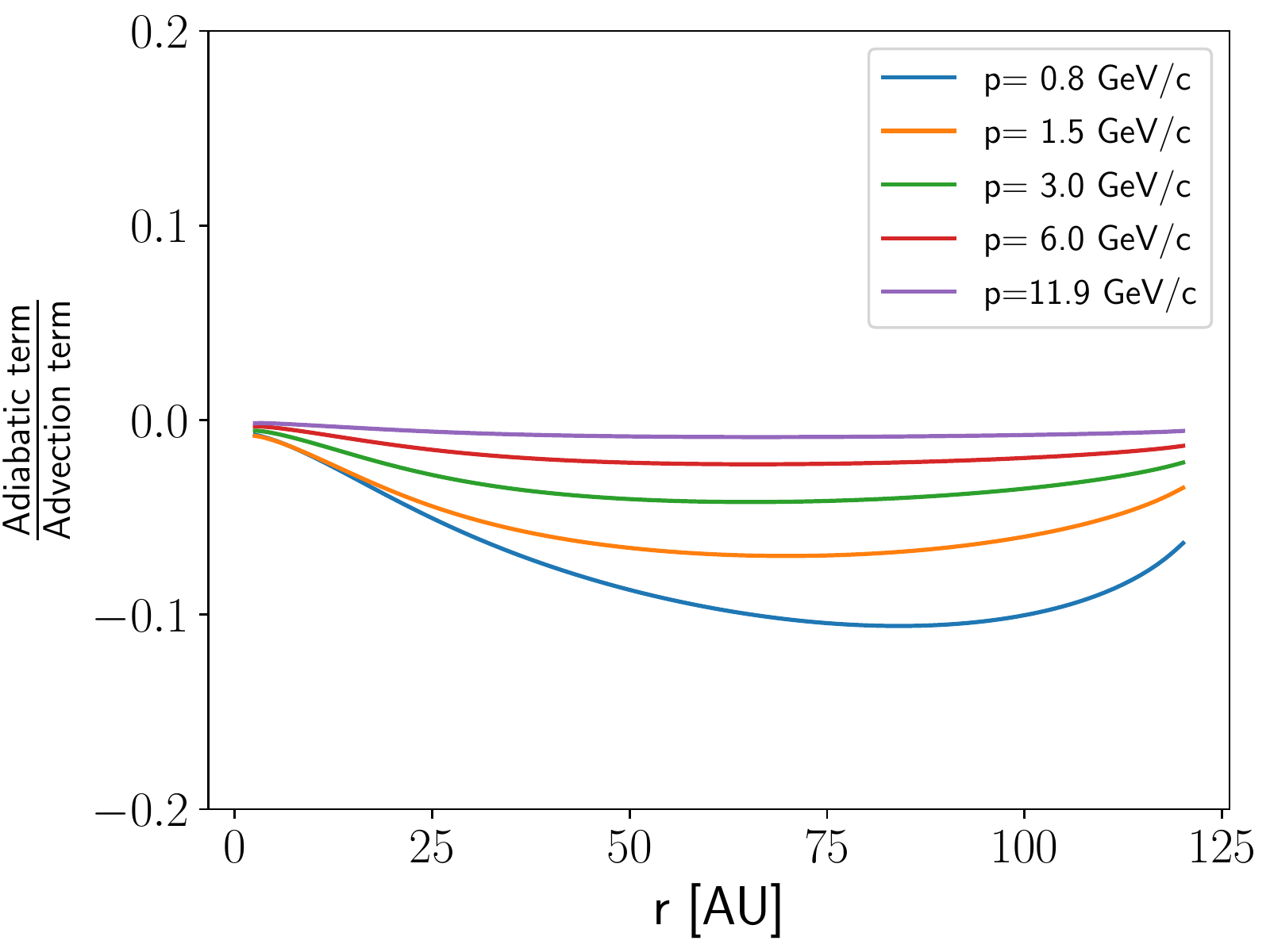}
\caption{Ratio of the adiabatic energy loss in eq.~\eqref{eqn:transport} to the advection term for different rigidities as a function of radial distance for a tilt angle of $\alpha = 10^\circ$. The adiabatic term contributes less than 10\% to the transport equation in the inner heliosphere for intermediate energies. It gets slightly more important for low energies. The left panel is for $qA<0$ and the right panel for $qA>0$.}
\label{fig:adiabatic}
\end{figure}

Given the fully numerical code and the heliospheric model described above, we can check the validity of the assumption made in deriving eq.~\eqref{eqn:semi-analytical_model1}. The most fundamental one is the neglect of the adiabatic term in the transport equation~\eqref{eqn:transport}. In Fig.~\ref{fig:adiabatic}, we compare the contribution from the adiabatic term, $\partial/\partial p \left( p^3 \vec{V} \cdot \nabla f \right) / (3 p^2)$ to the advective term, $\nabla \cdot ( C \vec{V} f )$. (The comparison of the adiabatic to the diffusive term $\nabla \cdot ( K \cdot \nabla f )$ leads to similar results.) It can be seen that the contribution from the adiabatic term is larger in the outer heliosphere than close to Earth. Furthermore, at rigidities above $\sim 1 \, \text{GV}$, the adiabatic term generally contributes less than $10 \, \%$ to the transport equation, in particular for $qA > 0$. We thus conclude that the neglect of the adiabatic term is justified.

\begin{figure}
\includegraphics[width=.5\textwidth]{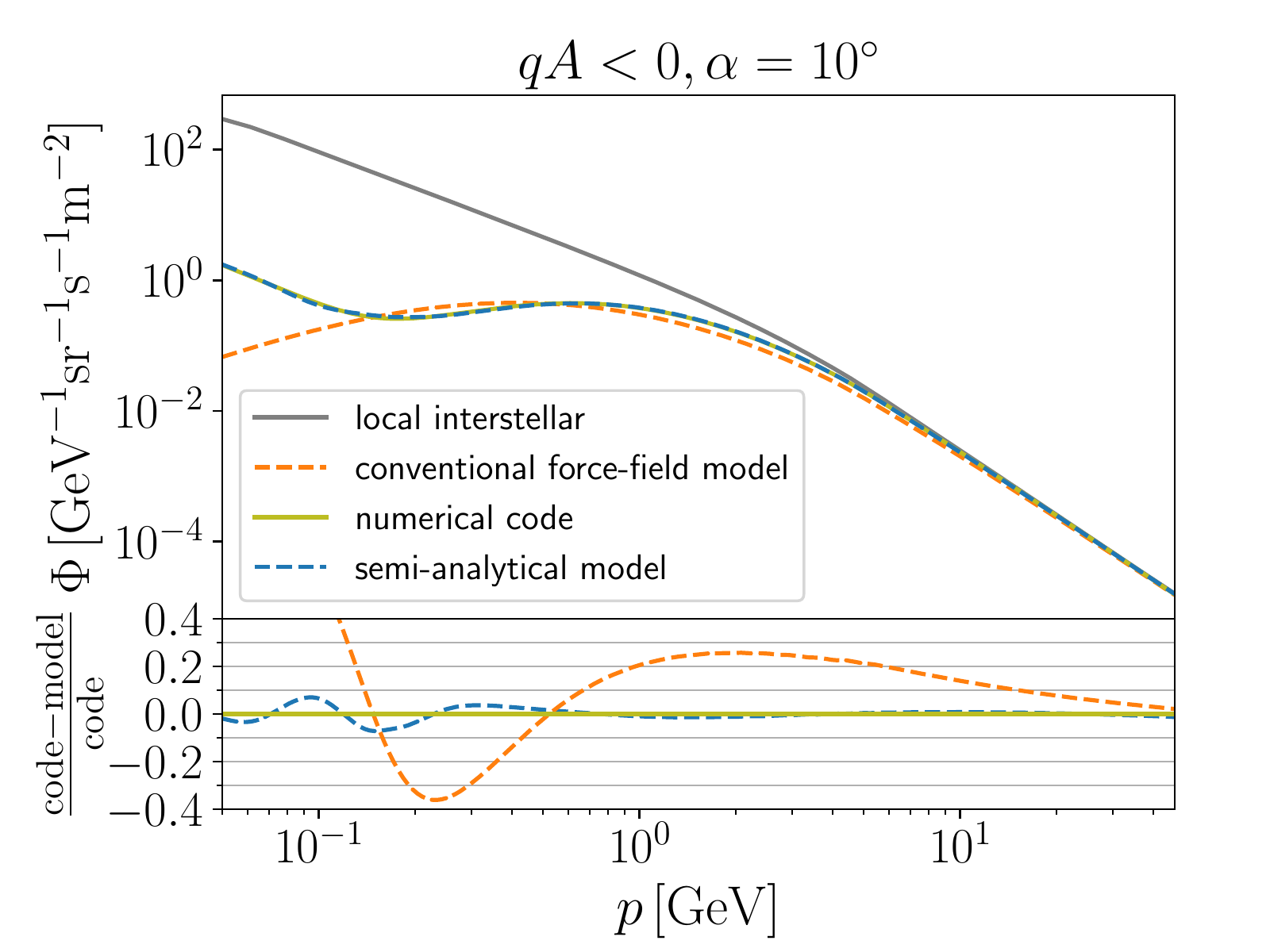}\includegraphics[width=.5\textwidth]{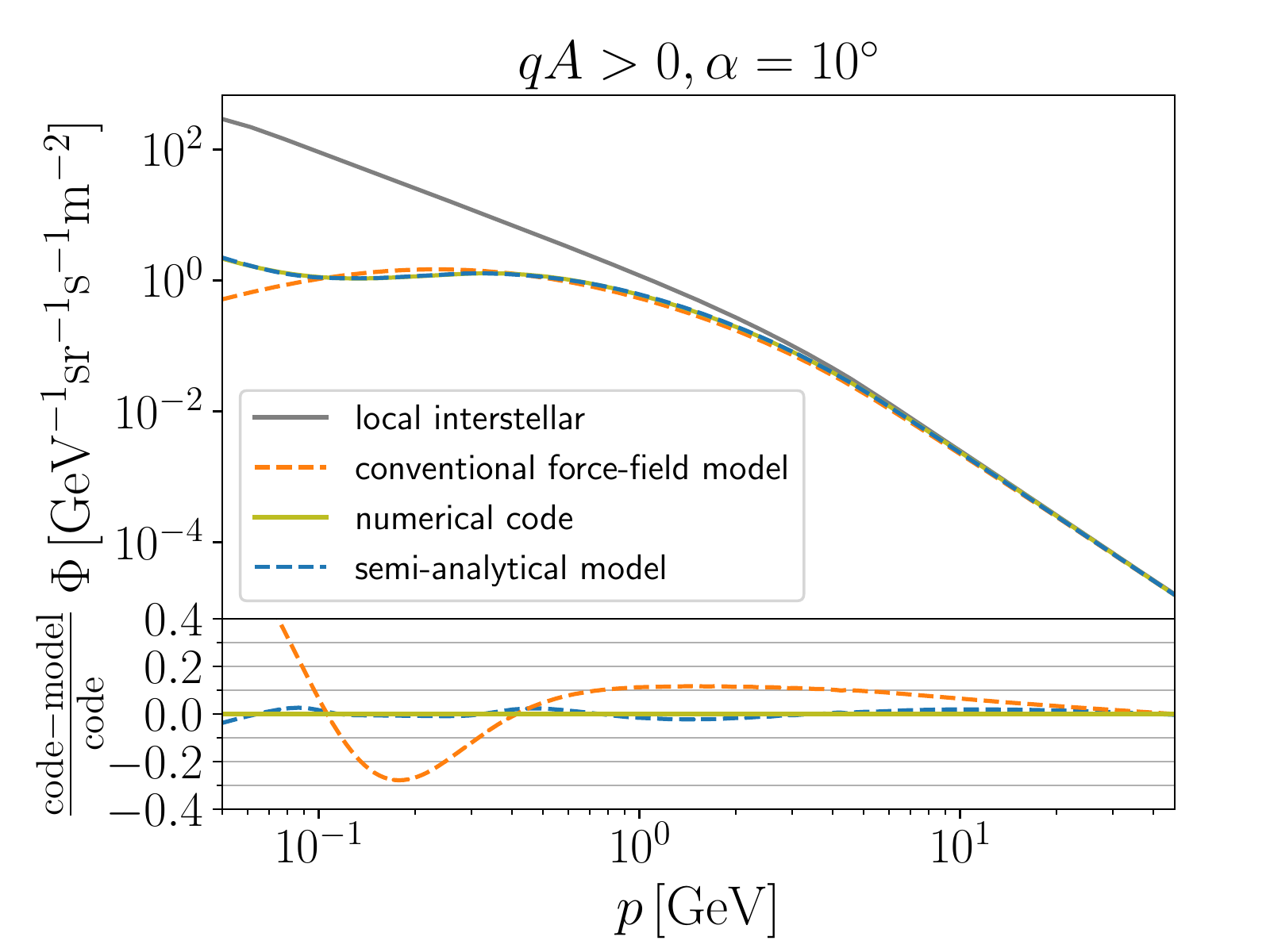}
  \caption{Fit of our semi-analytical model to the
  full numerical solution of the transport equation for a tilt angle of $\alpha
  = 10^\circ$. The left plot shows the fit for $qA<0$ and the plot on the right $qA>0$. In the lower panels the relative deviation from the numerical
  solution is shown.}
  \label{fig:codefit}
\end{figure}

Next, we check whether the fluxes modulated according to eqs.~\eqref{eqn:semi-analytical_model1} and \eqref{eqn:semi-analytical_model2} agree with the results from the fully numerical code. The transport parameters are as explained in the main part of the letter. In Fig.~\ref{fig:codefit} we show the results of fitting our semi-analytical model (eqs.~\ref{eqn:semi-analytical_model1} and \ref{eqn:semi-analytical_model2}) to the results from the numerical code. The fit agrees with the numerical solution to within less than 10\%. We find significantly better agreement than with the conventional force-field model.

\section{Cookbook for semi-analytic solar modulation}

\subsection{Solar modulation for individual times}

To reproduce our results for fitting to data for individual time bins follow this recipe:
\begin{enumerate}
  \item Chose parametrisations for the diffusion coefficient and the drift term.
The diffusion coefficient can for example be modelled as a softly broken power law in rigidity $R$ (understood to be measured in GV),
\begin{equation}
  \nonumber
\tilde{K}_{rr} = K_0R^a\left(\frac{R^c+R_k^c}{1+R_k^c}\right)^{(b-a)/c} \, ,
\end{equation}
with a normalization $K_0 = 30\,\text{AU}^2/\text{d}$, power law indices $a = 0$ and $b = 1.55$, $c=3.5$ and a break rigidity $R_k = 0.28$. An example parametrisation for the radial drift velocity is
\begin{equation}
  \nonumber
\tilde{v}_r = \kappa_{\text{0}} \frac{ R}{3 B_0}\frac{10\,R^2}{1+10\,R^2} \, ,
\end{equation}
where we set $\kappa_\text{0} = 1\,\text{AU}/\text{d}$ and the magnetic field $B_0 = 1 \, \mu\text{G}$.
\item Calculate the modulated phase-space density using
\begin{equation}
\tilde{f}(r, p) \! = \! f_{\mathrm{LIS}}(p_{\mathrm{LIS}}(r, p)) \exp \left[ - \int_0^r \mathrm{d} r' g_2\frac{\tilde{v}_r(r', p_{\mathrm{LIS}}(r', p))}{\tilde{K}_{r r}(r', p_{\mathrm{LIS}}(r', p)} \right] \,
\label{eqn:modulate}
\end{equation}
where we have defined $p_\text{LIS}(r',p') = p_{r'\!,p'}(r_\text{max})$ with $p_{r'\!,p'}(r)$ a solution of the initial value problem
\begin{equation}
\frac{\mathrm{d} p}{\mathrm{d} r} = g_1\frac{p \tilde{V}}{3 \tilde{K}_{r r}} \, ,
\end{equation}
for given transport parameters from an unmodulated flux as a function of the scaling factors $g_1$ and $g_2$. (You may use the example script provided at {\protect\url{https://git.rwth-aachen.de/kuhlenmarco/effmod-code}}).
\item Determine the scaling factors $g_1$ and $g_2$ by fitting.
\end{enumerate}

\subsection{Time dependent solar modulation}

To reproduce our results for the time dependent fluxes using a linear model of the scaling factors:

\begin{enumerate}
  \item Get magnetic field data from ACE (\protect\url{http://www.srl.caltech.edu/ACE/ASC/DATA/level3/mag/ACESpec.cgi?LATEST=1}) and tilt angle data from WSO (\protect\url{http://wso.stanford.edu/Tilts.html}).
  \item Calculate moving averages of the solar wind parameters using the window widths given in Tbl.~1 of the main paper.
  \item Calculate the scaling factors from the averaged solar wind parameters using $g_1^{\pm} = a_1^\pm + b_1^\pm \langle \alpha \rangle_{\tau}$ and $g_2^{\pm} = a_2^\pm + b_2^\pm \langle B \rangle_{\tau}$ with the coefficients from Tbl.~1 of the main paper.
  \item Calculate the modulated flux from the unmodulated flux and the scaling factors using equation \eqref{eqn:modulate}. (You may use the example script provided at {\protect\url{https://git.rwth-aachen.de/kuhlenmarco/effmod-code}}).
\end{enumerate}


\bibliographystyle{unsrtnat}
\bibliography{Bibliography}